# A topological quantum optics interface

Sabyasachi Barik[1,2], Aziz Karasahin[3], Chris Flower[1,2], Tao Cai[3], Hirokazu Miyake[2], Wade DeGottardi[1,2], Mohammad Hafezi[1,2,3]*, Edo Waks[2,3]*

[1]Department of Physics, University of Maryland, College Park, MD 20742, USA

[2]Joint Quantum Institute, University of Maryland, College Park, MD 20742, USA and National Institute of Standards and Technology, Gaithersburg, MD 20899, USA

[3]Department of Electrical and Computer Engineering and Institute for Research in Electronics and Applied Physics, University of Maryland, College Park, Maryland 20742, USA

*Correspondence to: hafezi@umd.edu, edowaks@umd.edu.

**Abstract**: The application of topology in optics has led to a new paradigm in developing photonic devices with robust properties against disorder. Although significant progress on topological phenomena has been achieved in the classical domain, the realization of strong light-matter coupling in the quantum domain remains unexplored. We demonstrate a strong interface between single quantum emitters and topological photonic states. Our approach creates robust counter-propagating edge states at the boundary of two distinct topological photonic crystals. We demonstrate the chiral emission of a quantum emitter into these modes and establish their robustness against sharp bends. This approach may enable the development of quantum optics devices with built-in protection, with potential applications in quantum simulation and sensing.

The discovery of electronic quantum Hall effects has inspired remarkable developments of similar topological phenomena in a multitude of platforms ranging from ultracold neutral atoms (1,2), to photonics (3,4) and mechanical structures (5-7). Like their electronic analogs, topological photonic states are unique in their directional transport and reflectionless propagation along the interface of two topologically distinct regions. Such robustness has been demonstrated in various electromagnetic systems, ranging from microwave domain (8,9) to the optical domain (10,11), opening avenues for a plethora of applications, such as robust delay lines, slow-light optical buffers (12), and topological lasers (13-15), to develop optical devices with built-in protection. While the scope of previous works remains in the classical electromagnetic regime, a great deal of interesting physics could emerge by bringing topological photonics to the quantum domain. Specifically, integrating quantum emitters to topological photonics structures could lead to robust strong light-matter interaction (16), generation of novel states of light and exotic many-body states (17-19).

In this work, we experimentally demonstrate light-matter coupling in a topological photonic crystal. We utilize an all-dielectric structure (20,21) to implement topologically robust edge states at the interface between two topologically distinct photonic materials, where the light is transversally trapped in a small area, up to half of the wavelength of light. We show that a quantum emitter efficiently couples to these edge modes and the emitted single photons exhibit robust transport, even in the presence of a bend. Figure 1A shows an image of the topological photonic structure which is comprised of two deformed honeycomb photonic crystal lattices made of equilateral triangular air holes on a GaAs membrane (21,22). Fig. 1B shows a closeup image of the interface, where the black dashed lines identify a single unit cell of each photonic crystal. In each region, we perturb the unit cell by concentrically moving the triangular holes either inward (yellow region) or outward (blue region). Fig. 1C-D shows the corresponding band structures of the two regions. The perturbations open two bandgaps exhibiting band inversion at the Γ point (20,21). Specifically, the region with compressed unit cell, highlighted in yellow, acquires a topologically trivial band gap, while the expanded region, highlighted in blue, takes on a nontrivial one. We design both regions so that their bandgaps overlap. Photons within the common bandgap cannot propagate into either photonic crystal. However, because the crystals have different topological band properties, the interface between them supports two topological helical edge modes, travelling in opposite directions, with opposite circular polarizations at the center of the unit cell.

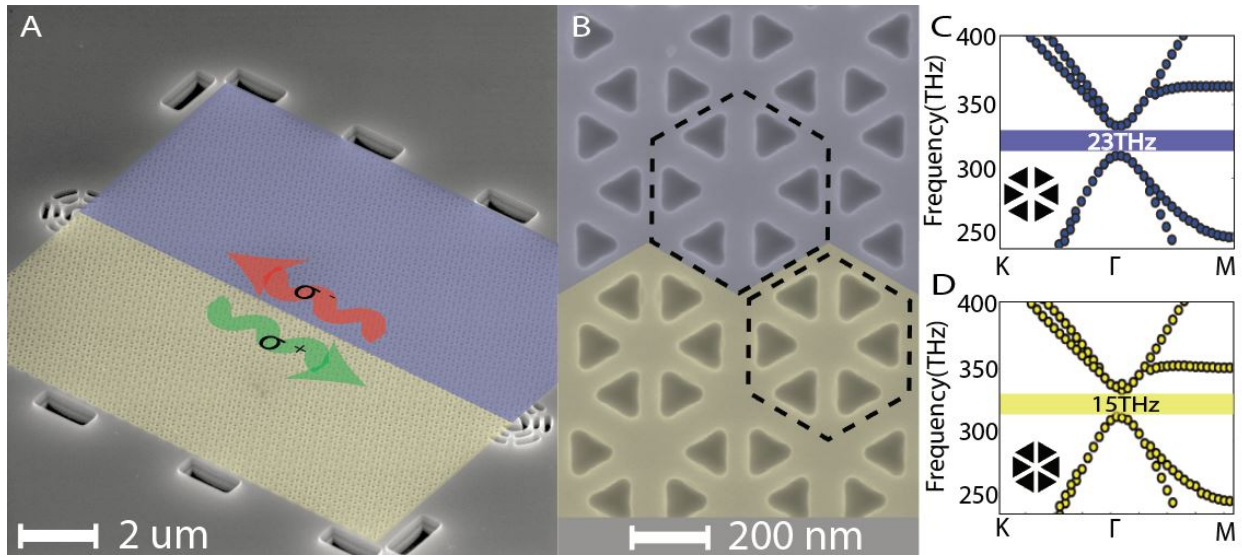

**Fig.1** : **Fabricated device and band structure. (A)** Scanning electron microscope image of the device composed of two regions identified by blue and yellow highlights, corresponding to two photonic crystals of different topological properties. The interface between the two photonic crystals supports helical edge states with opposite circular polarization ($\sigma^+/\sigma^-$). Grating couplers at each end of the device scatter light in the out-of-plane direction for collection. **(B)** Closeup image of the interface. **(C)** and **(D)** show the band structures for the transverse electric modes of the two photonic crystals.

To show the presence of the guided edge mode, we measure the transmission spectrum. We illuminate the left grating (L) with a 780 nm continuous-wave laser using a pump power of 1.3μW, and collect the emission from the right grating (R) (see Fig.2A). At this power the quantum dot ensemble emission becomes a broad continuum due to power broadening, resulting in an internal white light source that spans the wavelength range of 900-980 nm. Fig. 2B shows the spectrum at the right grating, presented with the band structure simulation (21). Light emitted within the topological band efficiently transmits through the edge mode and propagates to the other grating coupler, while photons outside of the bandgap dissipate into bulk modes.

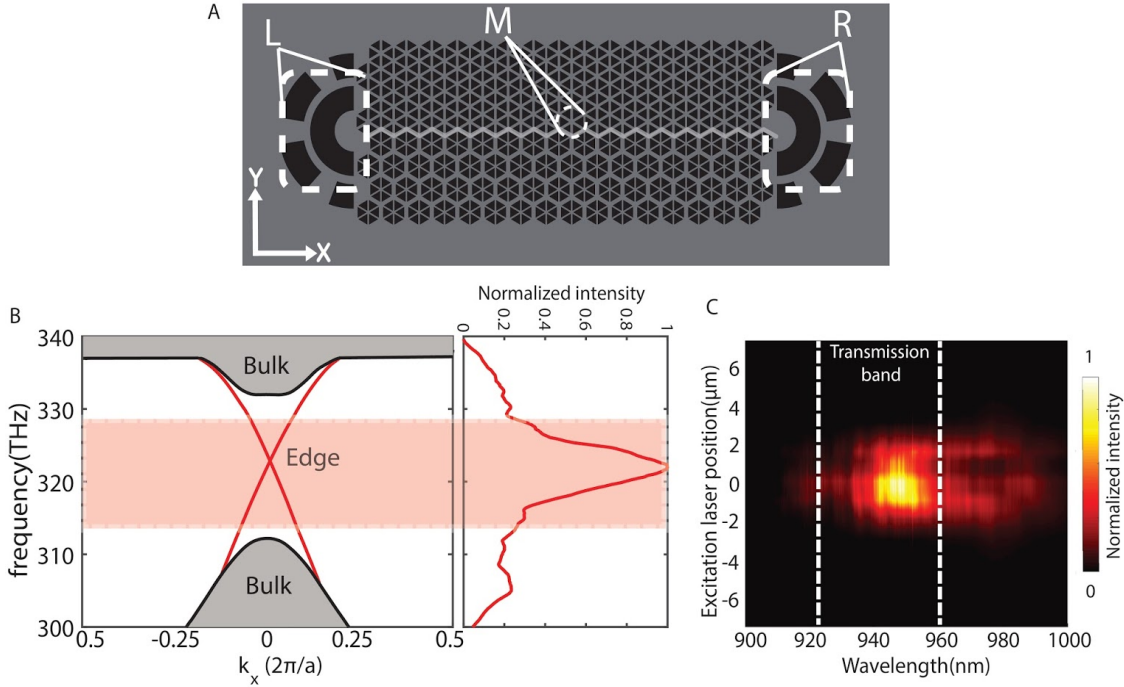

**Fig.2** : **Transmission characteristics of the topological waveguide. (A)** A schematic of the excitation scheme identifying the three relevant regions. **(B)** Simulated band structure of transverse electromagnetic modes of a straight topological waveguide. The grey region corresponds to bulk modes of the individual topological photonic crystals and red lines represent modes within the bandgap corresponding to topological edge states. The adjacent panel shows the measured spectrum at the transmitted end of the waveguide. The shaded region identifies the topological edge band. **(C)** Transmission spectrum at grating L as a function of the excitation laser position.

To confirm that the emission originates from guided modes at the interface between the two topological materials, we excite the structure in the middle of the waveguide (M), and collect the emission at the left grating coupler. Fig. 2C shows the transmission spectrum as a function of the laser spot position as we scan the laser along the *y*-axis across the interface. The spectrum attains a maximum transmission within the topological band when the pump excites the center of the structure. When we displace the beam, by approximately 1.5 microns, the transmission vanishes, indicating that the photons are coming only from the waveguide.

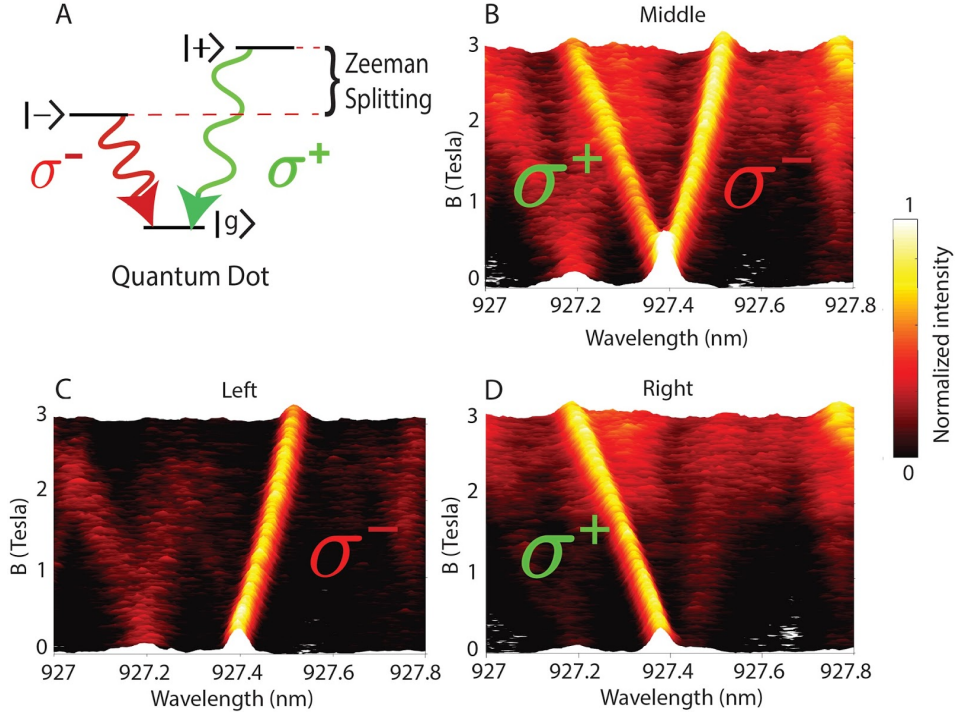

**Fig. 3**: **Chirality in a straight topological waveguide.** (**A**) Schematic of quantum dot level structure in the presence of a magnetic field, and radiative transitions with opposite circular polarizations. (**B**) Emission spectrum collected from the excitation region as a function of magnetic field. (**C**) and (**D**) Transmission spectrum to left and right gratings, respectively.

A key feature of topological edge modes is the chiral nature of the coupling between the helical topological edge mode and the quantum emitter. Specifically, different dipole spins radiatively couple to opposite propagating helical edge states. To demonstrate this helical light-matter coupling, we apply a magnetic field in the out-of-plane (Faraday) direction. This field induces a Zeeman splitting in the quantum dot excited state, resulting in two non-degenerate states with total angular momentum of ±1 (Figure 3A) (23). These states emit with opposite circular polarization to the ground state via the optical transitions denoted as $\sigma^\pm$. By spectrally resolving the emissions we can identify the dipole spin and correlate it with the propagation direction of the emitted photon.

To isolate a single quantum emitter within the topological edge mode, we reduce the power to 10 nW, which is well below the quantum dot saturation power. Using the intensities of the collected light at the two ends, we calculate a lower bound on the collection efficiency of 68%, defined as the ratio of the photon emission rate into the waveguide to the total emission rate (22). This high efficiency is due to the tight electromagnetic confinement of the guided modes which enhances light-matter interactions. Figure 3B shows the emission spectrum as a function of magnetic field. As the magnetic field increases, the quantum dot resonance splits into two branches corresponding to the two Zeeman split bright exciton states. We compare this spectrum to the one collected from left and right gratings (Fig. 3 C-D). At the left grating we observe only the emission from the σ⁻ branch, while at the right grating we observe only the σ⁺ branch. These results establish the chiral emission and spin-momentum locking of the emitted photons, and provide strong evidence that the emitter is coupling to topological edge states that exhibit unidirectional transport. Such chiral coupling is in direct analogy to one dimensional systems (24,25,16). In contrast to one dimensional systems, the waveguided modes of our structure originate from two dimensional topology. As a result, the topological edge mode should exhibit robustness to certain deformations, such as bends.

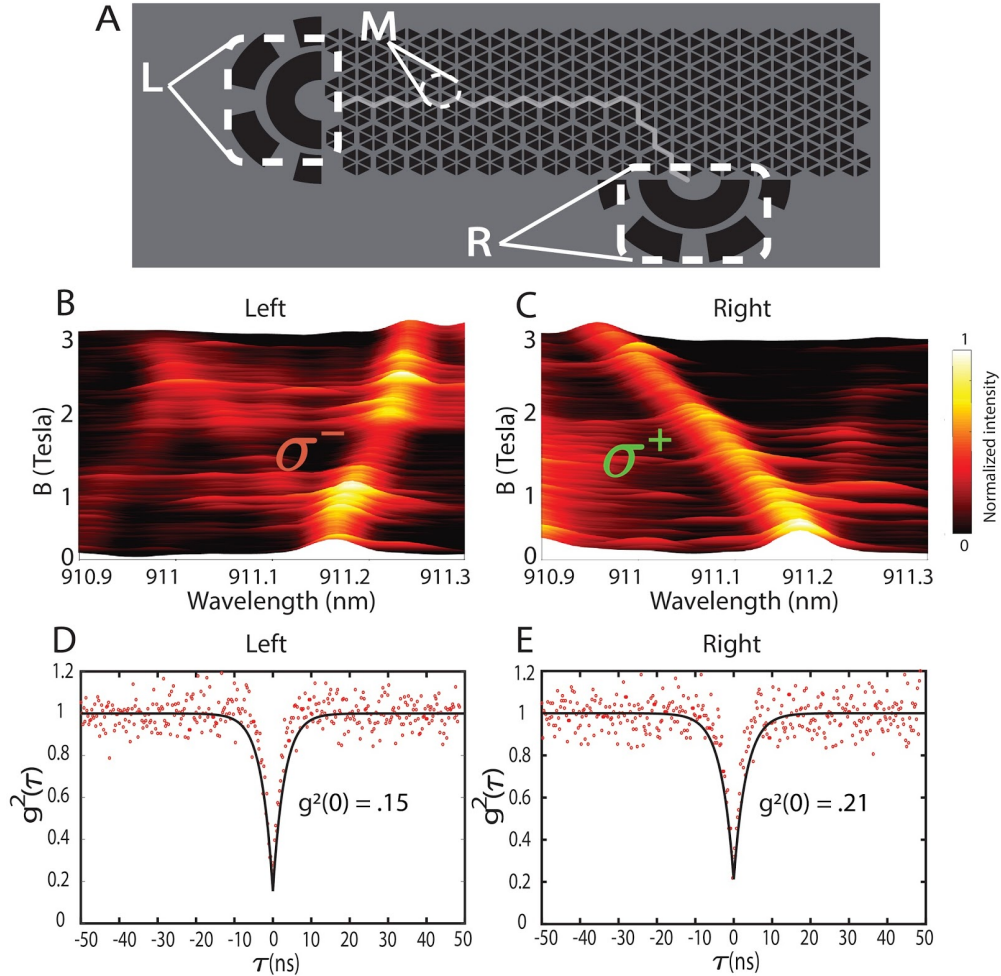

**Fig. 4 : Robust transport in two dimensions along a bend:** **(A)** Schematic of a modified topological waveguide with a bend. **(B)** and **(C)** Photoluminescence collected from position L and position R, respectively, showing only one branch of the quantum dot. **(D)** and **(E)** Second-order correlation measurement ($g^2(\tau)$) data obtained from point L and R, respectively, showing anti-bunching.

In order to establish this topological robustness, we analyze the propagation of emitted photons in the presence of a bend. We introduce a 60 degree bend into the structure as shown in Fig. 4A, and perform measurements similar to those in Fig. 3. Again we observe that emitted photons propagate in opposite direction in a chiral fashion and arrive at the grating associated with their respective polarization(Fig. 4 B-C). The preservation of the chiral nature of the emission demonstrates an absence of back-reflection at the bend, which would result in a strong signal for both polarizations at the left grating. We also confirm that these routed photons are indeed single photons by performing a second order correlation measurement for photons collected from both ends of the waveguide, which exhibits strong anti-bunching (Fig. 4 D-E).

In this work, we demonstrated coupling between single quantum emitters and topologically robust photonic edge states. The present approach opens up new prospects at the interface of quantum optics and topological photonics. In the context of chiral quantum optics, one can explore new regimes of dipole emission in the vicinity of a topological photonic structures and exploit the robustness of the electromagnetic modes (16). Furthermore, in a chiral waveguide, photon-mediated interactions between emitters are location-independent (26). This property could facilitate the coupling of multiple solid-state emitters via photons while overcoming scalability issues associated with random emitter position, enabling large-scale super-radiant states and spin-squeezing. Ultimately, such an approach could form a versatile platform to explore many-body quantum physics at a topological edge (27), create chiral spin networks (26,28), and realize fractional quantum Hall states of light (29,30).

**Acknowledgments:** The authors would like to acknowledge fruitful discussions with Sunil Mittal. This research was supported by ONR, AFOSR, Sloan Foundation and the Physics Frontier Center at the Joint Quantum Institute.

## SUPPLEMENTARY MATERIALS:

## Materials and Methods

### Device Design

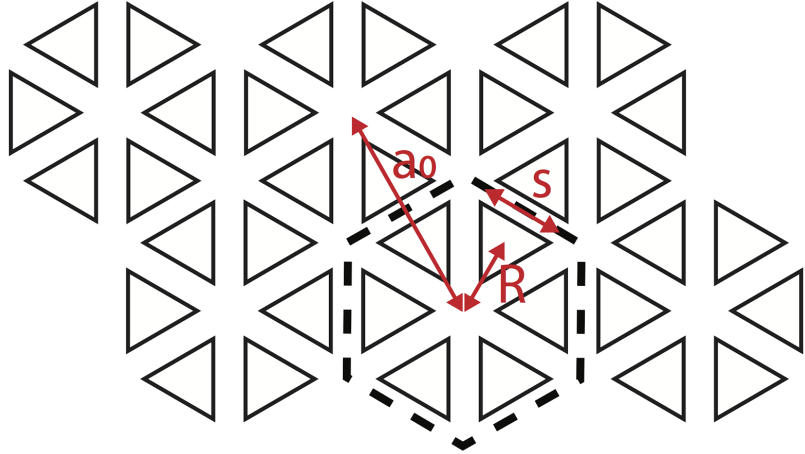

**Figure S1: Design of honeycomb-like photonic crystal.**

Figure S1 shows a schematic of the device design. We begin with a honeycomb lattice of equilateral triangles exhibiting hexagonal symmetry as our baseline structure. This lattice is a triangular lattice of cells consisting of six equilateral triangular holes, indicated by the dashed line. We use a lattice constant of $a_0 = 445$ nm, an edge length of the equilateral triangle of $s=140$ nm, and a slab thickness of $h = 160$ nm. $R$ defines the distance from the center of a cell to the centroid of a triangle. In this structure a perfect honeycomb lattice corresponds to $R = a_0/3$.

With these parameters we obtain doubly degenerate Dirac cones at 319 THz (940 nm). We form the two mirrors by concentrically expanding or contracting the unit cell.

We create topologically distinct regions by deforming the unit cell of the pristine honeycomb lattice. In the blue region in Fig. 1A, we concentrically shift the triangular holes by increasing $R$ to $1.05a_0/3$, thereby shifting all the triangular holes in an individual cell outward. This deformation results in the band structure shown in Figure 1C. In the yellow region, we decrease $R$ to $0.94a_0/3$, which pulls the holes towards the center resulting in the band structure shown in Figure 1D.

**Device Fabrication:**

To fabricate the device, we began with an initial wafer composed of a 160 nm GaAs membrane on top of 1 $\mu$m sacrificial layer of $Al_{0.8}Ga_{0.2}As$ with quantum dots grown at the center. The quantum dot density was approximately 50 $\mu m^{-2}$. We fabricated the topological photonic crystal structure using electron beam lithography, followed by dry etching and selective wet etching of the sacrificial layer. We first spin-coated the wafer with ZEP520A e-beam resist, then patterned the structure using 100 keV acceleration voltage and developed the resist using ZED50 developer. After patterning, we used chlorine-based inductively coupled plasma etching to transfer the pattern on the GaAs membrane. We finally performed selective wet etching using HF to create a suspended structure with air on top and bottom. The rectangular structures in the periphery are included to facilitate undercut of the sacrificial layer.

Sharp corners with straight side walls are essential to observe the topological chiral edge modes. It is confirmed via simulation that triangles with rounded corners are detrimental for the device operation. However, even with highly directional dry etch, creating sharp features like triangles is challenging at such small length scales.

We observed – by using a regular mask design (as shown in Figure S2.A) – that etching causes widening of holes which eventually results in rounded corners much like a Reuleaux triangle (Figure S2.B). We used a modified mask design to overcome this challenge. Triangles with shrunk edges shown in Figure S2.C are used as a mask; this results in sharp triangles with edge lengths of 140 nm. Close up SEM image of final structure is shown in Figure S2.D.

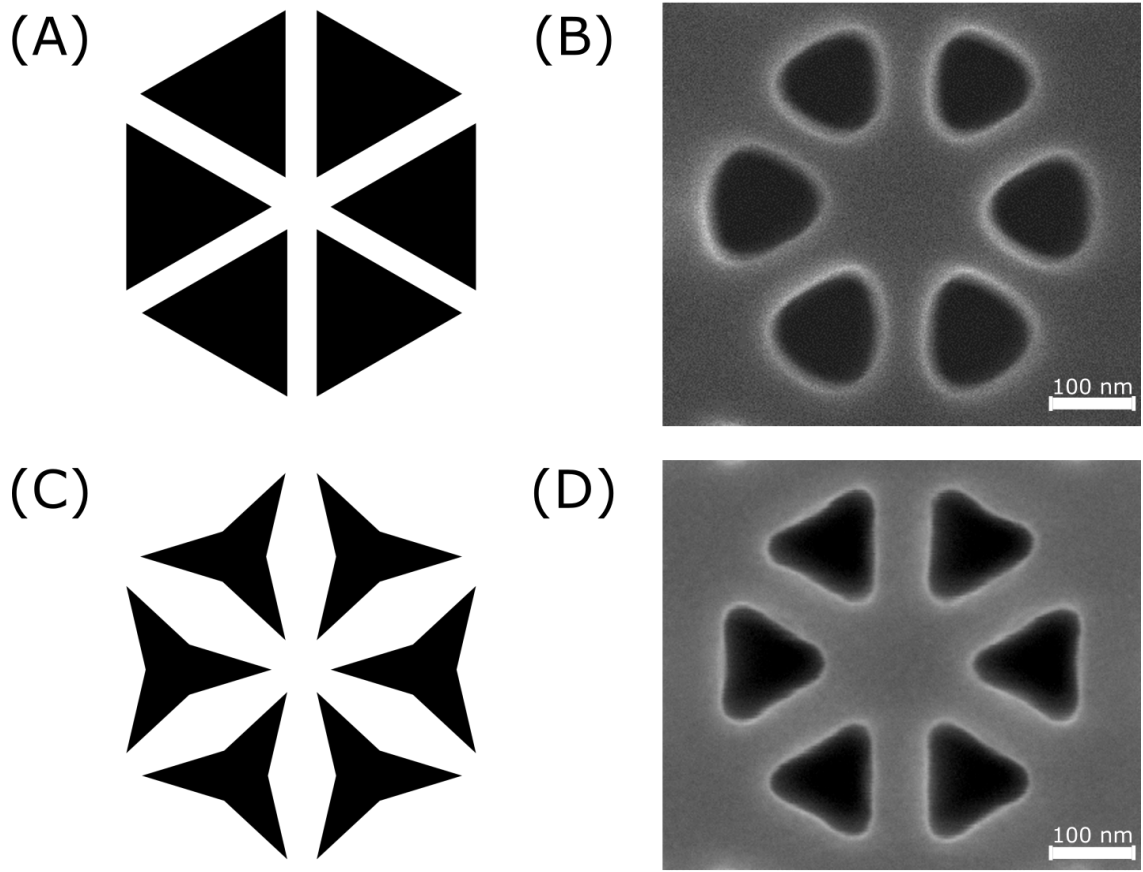

**Figure S2: Mask design for fabrication of triangles:** (**A**) Layout of regular mask. (**B**) SEM image of rounded triangles resulted from use of regular mask. (**C**) Layout of modified mask; triangles are bent from edges to mitigate etching imperfections. (**D**) SEM image of sharp triangles fabricated with use of modified mask.

**Coupling efficiency**

The coupling efficiency of emission from a single quantum emitter into the topological waveguide is defined by $\beta = \frac{I_L + I_R}{I_L + I_R + I_M}$, where $I_i$ ($i = L, R, M$) is the integrated photon counts from left, right or middle of the waveguide, respectively. Since we do not have a reliable estimate of the grating coupler efficiencies, this figure gives a lower bound on the ratio of the photon emission rate into the waveguide to the total emission rate.

**Experimental setup**

To perform measurements, we mounted the sample in a closed-cycle cryostat and cooled it down to 3.6 K. A superconducting magnet applied a magnetic field of up to 9.2 T along the out-of-plane (Faraday) direction in order to generate a Zeeman splitting between the two bright excitons of the quantum dot. We performed all sample excitation and collection using a confocal microscope with an objective lens with numerical aperture of 0.8. We collected the emission and focused it onto a single mode fiber to perform spatial filtering. To perform spectral measurements, we injected the signal to a grating spectrometer with a spectral resolution of 7 GHz. For autocorrelation measurements, we used a flip mirror to couple the light out of the spectrometer and processed the filtered emission using Hanbury-Brown Twiss intensity interferometer composed of a 50/50 beamsplitter, two Single Photon Counting Modules (SPCMs) and a PicoHarp 300 time correlated single photon counting system.